\documentclass[twocolumn,showpacs]{revtex4}

\usepackage{bm}

\usepackage{graphicx}
\newcommand{\simle}
{\raisebox{-0.75ex}[-1.5ex]{$\;\stackrel{<}{\sim}\;$}}
\newcommand{\simge}
{\raisebox{-0.75ex}[-1.5ex]{$\;\stackrel{>}{\sim}\;$}}

\usepackage{amssymb}

\begin{document}

\title{A Microscopic Theory of Odd-freqeuncy Pairing in the Two-dimensional Extended Hubbard Model}

\author{Keiji Yada}
\affiliation{Toyota Physical and Chemical Research Institute, Nagakute-cho 480-1192, Japan}
\author{Seiichiro Onari}
\author{Yukio Tanaka}
\affiliation{Department of Applied Physics, Nagoya University, Nagoya 464-8603, Japan}
\author{Kazumasa Miyake}
\affiliation{Department of Physical Science, Osaka University, Osaka 560-8531, Japan}

\begin{abstract}
We present a microscopic theory of fluctuation-mediated 
pairing mechanism in a two-dimensional extended Hubbard model.
In contrast to conventional wisdom, odd-frequency spin-triplet pairing
can be stabilized near the spin-density-wave critical point. 
Favorable conditions for the odd-frequency pairing 
are the presence of geometrical frustration and 
off-site Coulomb interaction. 
\end{abstract}


\maketitle
Exploring unconventional pairing functions of superconductivity 
in strongly correlated electron systems 
has been an important issue in condensed matter physics. 
The pairing function of two electrons 
must change sign under the exchange of two electrons 
in accordance with Fermi-Dirac statistics.
Usually the pairing function is an even (odd) function of the momentum 
when the spin state of the pair is a spin-singlet (-triplet),
since we postulate that the pairing function is even as a function of frequency. However, 
the pairing function can be odd, as a function of the frequency 
where the resulting pairing function has odd (even) parity in momentum space 
for the spin-singlet (-triplet) pairing. 
For example, the "spin-triplet $s$-wave" or the "spin-singlet $p$-wave" with "odd frequency" can emerge. \par

Odd frequency superconductivity has been discussed by Berezinskii for the
$s$-wave spin-triplet case\cite{bere} in the context of $^{3}$He. 
Subsequently, there have been several theoretical studies
on odd-frequency pairings \cite{bala,Vojta,Bulut,Coleman}. 
Fuseya, {\it et al}. have discussed the possibility of "spin-singlet $p$-wave" pairing
near the antiferromagnetic (AF) quantum critical point and in the coexisting region with AF
by using phenomenological pairing interactions in the context of the superconducting state in Ce compounds \cite{Fuseya,Zhang}.
However, a microscopic mechanism to realize odd-frequency pairing 
has not been established yet, since it is not easy to imagine 
superconductors without equal time pairing. 
\par

Although realization of the superconducting state, $i.e.$, 
pair potential in the bulk system, has not been clarified, 
odd-frequency pairing correlation, $i.e.$, pair amplitude, 
is revealed to be generated in inhomogeneous systems 
where the bulk superconductor is a conventional even-frequency one. 
In ferromagnet/superconductor heterostructures with inhomogeneous
magnetization,
the odd-frequency pairing state was first proposed in
Ref. \onlinecite{Efetov1} and 
various aspects of this state were subsequently intensively
studied \cite{Efetov2}. 
Furthermore, the ubiquitous presence of
odd-frequency pairing amplitude in 
non-uniform superconducting systems has been revealed 
even if the bulk superconductor 
is a conventional even-frequency one \cite{triplet,inhomo}. 
Nowadays, the physics of 
odd-frequency pairing has become an exciting topic for study. 

On the basis of this background, 
it is an intriguing issue to clarify 
whether the odd-frequency bulk superconducting 
state can be derived from a microscopic model in a 
strongly correlated system. 
It is well known that the spin-singlet 
$d_{x^2-y^2}$-wave state is favored due to AF fluctuations
in the Hubbard Model in a square lattice
\cite{Miyake}.
On the other hand, the spin-singlet $d$-wave state
is not favored in a triangular lattice 
due to geometrical frustration\cite{Kondo}, wherein the 
nesting condition becomes worse.
Vojta, {\it et al}. discussed
the possibility of odd-frequency pairing 
in a triangular lattice\cite{Vojta}. 
The merit of the odd-frequency pairing is
that it can avoid direct Coulomb repulsion 
since pairs are formed by two electrons at
different times by the retardation effect of 
interaction between two electrons. 
It is highly desirable to clarify favorable conditions 
which generate odd-frequency pairing 
on the basis of a microscopic calculation. \par
The purpose of the present study is to explore the odd-frequency pairing state 
in the extended Hubbard model on a 2D triangular lattice within the random phase approximation (RPA), 
wherein off-site Coulomb interaction $V$ as well as on-site Coulomb interaction $U$ exist. 
Near the spin-density-wave (SDW) states, 
odd-frequency pairing becomes dominant. 
In the presence of $V$, even-frequency spin-singlet $d_{x^2-y^2}$-wave pairing, in which two electrons on neighboring sites form pairs, is unfavorable
since $V$ suppresses the attractive interaction between them.
By changing the lattice structure from triangular to square, 
even-frequency spin-singlet $d_{x^2-y^2}$-wave pairing 
becomes dominant. 
Based on these results, preferable conditions for the realization of the 
odd-frequency pairing are 
i) geometrical frustration, ii) the presence of $V$, and 
iii) sufficiently large amount of $U$ which induces the SDW order.

We consider the extended Hubbard model at half-filling on a two-dimensional triangular lattice,
where on-site repulsive interaction $U$ and off-site interaction 
between nearest neighbor sites $V$ are included.
The Hamiltonian is given as
\begin{eqnarray}
H&=&\sum_{\langle i,j\rangle,\sigma}\left(t_{ij}c_{i\sigma}^\dag c_{j\sigma}+h.c.\right)
\nonumber\\&&{}+\sum_{i}Un_{i\uparrow}n_{i\downarrow}
+\sum_{\langle i,j\rangle}V_{ij}n_in_j,
\end{eqnarray}
where $c_{i\sigma}(^\dag)$ is an annihilation (a creation) operator of electron,$n_{i\sigma}=c_{i\sigma}^\dag c_{i\sigma}$ and $n_i=n_{i\uparrow}+n_{i\downarrow}$.
Within the RPA, the spin susceptibility $\chi_s$ and the charge susceptibility $\chi_c$ are given as
\begin{eqnarray}
\chi_s(q)&=&\frac{\chi_0(q)}{1-U\chi_0(q)},\\
\chi_c(q)&=&\frac{\chi_0(q)}{1+(U+2V({\bm q}))\chi_0(q)},
\end{eqnarray}
where $q\equiv({\bm q},i\omega_m)=({\bm q},2m\pi Ti)$.
Here $\chi_0$ is the irreducible susceptibility given by
\begin{eqnarray}
\chi_0(q)=-\frac{T}{N}\sum_{k}G(k)G(k+q),
\end{eqnarray}
where $k\equiv({\bm k},i\varepsilon_n)=({\bm k},(2n-1)\pi Ti$)
and $G(k)=(i\varepsilon_n-\varepsilon_{\bm k}+\mu)^{-1}$.
In the above, 
$n$ and $m$ are integers, $T$ denotes temperature and  $N$ denotes the number of sites. 
In a triangular lattice, the band dispersion $\varepsilon_{\bm k}$ and the off-site Coulomb interaction $V({\bm q})$ are given as
\begin{eqnarray}
\varepsilon_{\bm k}=-2t_{1}[\cos(k_x)+\cos(k_x)]-2t_{2}\cos(k_x+k_y),\\
V({\bm q})=2V(\cos(q_x)+\cos(q_y)+\cos(q_x+q_y)).
\end{eqnarray}
The effective pairing interactions for the spin-singlet and spin-triplet channels due to spin and charge fluctuations are given by
\begin{eqnarray}
V_{eff}^s(q)&=&U+V({\bm q})+\frac{3}{2}U^2\chi_s(q)
\nonumber\\&&{}-\frac{1}{2}(U+2V({\bm q}))^2\chi_c(q),\label{veffs}\\
V_{eff}^t(q)&=&V({\bm q})-\frac{1}{2}U^2\chi_s(q)
\nonumber\\&&{}-\frac{1}{2}(U+2V({\bm q}))^2\chi_c(q).
\end{eqnarray}
In order to study the pairing instabilities,
we solve the linearized Eliashberg equation with the power method,
\begin{equation}
\lambda\Delta(k)=-\frac{T}{N}\sum_{k'}V_{eff}(k-k')G(k')G(-k')\Delta(k').\label{gap}
\end{equation}
The superconducting transition temperature 
$T_{\rm c}$ is determined by the condition $\lambda=1$. 
We discuss the pairing instabilities by comparing the magnitude of
$\lambda$ for each pairing symmetry.
For the pairing function $\Delta(k)$,
we consider the following four symmetries:
(i) even-frequency spin-singlet even-parity  (ESE),
(ii) even-frequency spin-triplet odd-parity  (ETO),
(iii) odd-frequency spin-singlet odd-parity  (OSO),
(iv) odd-frequency spin-triplet even-parity (OTE).
In the present calculation, we choose $t_1=1$ for the unit of energy.
We set the number of electrons as $n=1$. 
We take $72\times72$ ${\bm k}$-meshes and 1024 Matsubara frequencies.

\begin{figure}[htbp]
\begin{center}
\includegraphics[height=6cm]{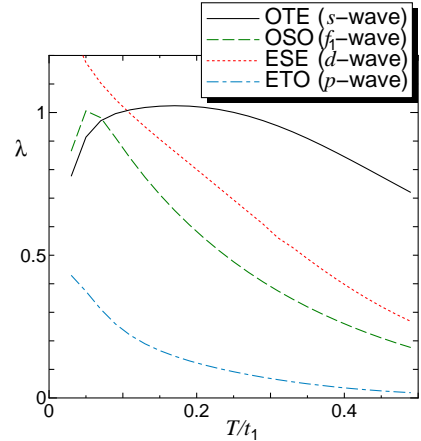}
\caption{$T$-dependence of $\lambda$ at $U/t_1=4.0$, $V/t_1=0$ in a regular triangular lattice.}\label{fig1}
\end{center}
\end{figure}

First, we focus on the isotropic triangular lattice with 
$t_{1}=t_{2}$.
In this case, the six irreducible representations of C$_{\rm 6v}$ group is possible for the pairing function.
In this paper, we use the notation $s$- $d$-, $i$-, $p$-, $f_1$- and $f_2$-wave for A$_1$, E$_2$, A$_2$, E$_1$, B$_1$ and B$_2$, respectively.
In Fig. \ref{fig1}, we show the $T$-dependence of $\lambda$ at half-filling with $U/t_1=4.0$ and $V/t_1=0$.
For each pairing symmetry (ESE, ETO, OSO and OTE),
we show the result with the largest eigenvalue $\lambda$. For the OSO pairing, the obtained values of $\lambda$ for the $p$-wave and $f_1$-wave are very close.
Thus, we only show the result for the $f_1$-wave. 
It is remarkable that the value of 
$\lambda$ for OTE pairing is the largest for 
a wide range of temperatures. 
This value has non-monotonic temperature dependence, 
where it reaches unity and then decreases with 
a decrease in $T$ for $T/t_1\simle0.1$.

In order to understand the 
generation of the OTE state, 
we analyze the momentum and frequency 
dependence of the pairing function. 
For this purpose, we transform Eq. (\ref{gap}) 
\begin{equation}
\lambda=-\frac{T}{N}
[\sum_{k,k'}f(k,k')]/[\sum_{k}|\Delta(k)|^2].
\label{gap2}
\end{equation}
with $f(k,k')=V_{eff}(k-k')G(k')G(-k')\Delta(k)\Delta(k')$.
First, we focus on the ESE pairing.
Note that there is no sign change in $\Delta(k)$ as a function of the Matsubara frequency for even-frequency pairing.
For spin-singlet pairing, the pairing interaction $V_{eff}(q)>0$ 
is satisfied when the spin fluctuations are dominant.
To obtain a large positive $\lambda$,
the pairing function $\Delta(k)$ 
should change sign by the shift of ${\bm Q}\sim(0.4\pi,0.4\pi)$, 
where $V_{eff}({\bm Q}, i\omega_m)$ takes the maximum value.
Therefore, nodes are formed in the pairing function consistent with 
$d_{x^{2}-y^{2}}$-wave pairing.

\begin{figure}[htbp]
\begin{center}
\includegraphics[height=6cm]{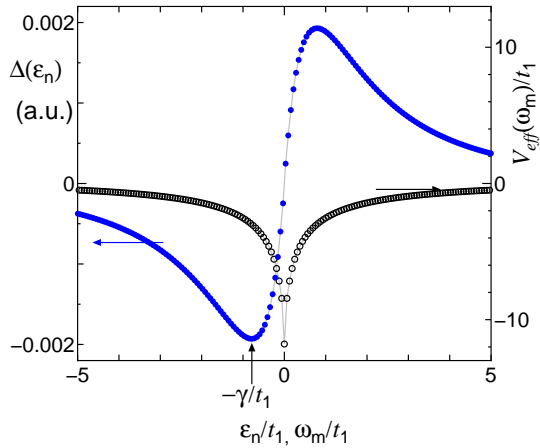}
\caption{Matsubara-frequency dependence of $\Delta({\bm k}_0,i\varepsilon_n)$ and $V_{eff}({\bm Q},i\omega_n)$
 for odd-frequency spin-triplet $s$-wave pairing for $U/t_1=3.5$ and $T/t_1=0.01$ .}\label{fig2}
\end{center}
\end{figure}

Next, we consider the OTE ($s$-wave). 
For spin-triplet pairing, $V_{eff}(q)$ always 
becomes negative.  
Therefore, the pairing function without nodes ($s$-wave) becomes dominant. 
In Fig. \ref{fig2}, we show the Matsubara frequency dependence of 
$\Delta({\bm k}_0, i\varepsilon_n)$ and $V_{eff}({\bm Q}, i\omega_n)$
for the OTE ($s$-wave) with $U/t_1=3.5$, $T/t_1=0.01$.
Here, 
${\bm k}_0=(\pi/2, \pi/2)$ denotes the wave vectors, where $\Delta({\bm k}_0, i\varepsilon_n)$ 
has the maximum value. 
To understand the 
frequency dependence, we ignore the momentum dependence of 
$V_{eff}(q)$ and $\Delta(k)$.
We introduce the half-value width of $V_{eff}(\omega_m)$, $\Gamma$,
and the position of peak $\gamma$ in $\Delta(\varepsilon_n)$. 
In the summation of the numerator on the right hand side of Eq. (\ref{gap2}),
$V_{eff}(\varepsilon_{n}-\varepsilon_{n'})$ for 
$|\varepsilon_{n}-\varepsilon_{n'}|<\Gamma$ mainly contributes to $\lambda$.
Since $\Delta(\varepsilon_n)$ is an odd-frequency of $\varepsilon_{n}$, 
$f(k,k')$ with $\varepsilon_{n}\varepsilon_{n'}>0$ enhances $\lambda$,
while that with $\varepsilon_{n}\varepsilon_{n'}<0$ suppresses $\lambda$.
In order to increase the value of $\lambda$,
$\gamma$ should be larger than $\Gamma$
because the negative parts of $f(k,k')$ 
are reduced with the decrease of
$|V_{eff}(\omega_m)|$ for $\omega_m>\Gamma$.
On the other hand, $G(k')G(-k')=1/(\varepsilon_{n'}^2+\varepsilon_{\bm k'}^2)$
has a sufficient value only for $|\varepsilon_{n'}|\simle T/t_1$.
If the magnitude of $\gamma$ is larger than $T/t_1$,
the positive contribution to enhance the magnitude of 
$\lambda$ becomes small 
due to the small amplitude of $\Delta(k')G(k')G(-k')$.
Therefore, the value of $\lambda$ for the OTE pairing 
decreases at low temperatures with $T/t_1\simle0.1$.
As a result, a large magnitude of 
$\lambda$ is realized for $T/t_1\simge \gamma\simge \Gamma$.

\begin{figure}[h]
\begin{center}
\includegraphics[height=6cm]{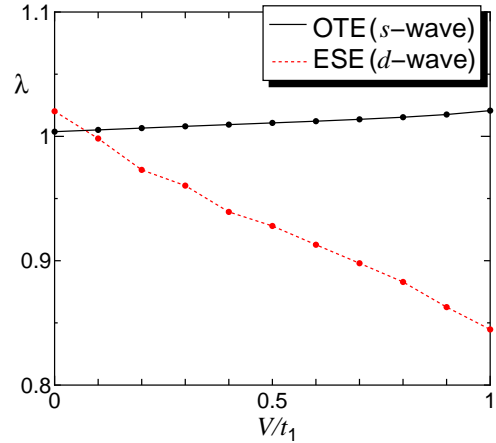}
\caption{$V$-dependence of $\lambda$ at $U/t_1=4.0$ and $T/t_1=0.1$.}\label{fig4}
\end{center}
\end{figure}

In Fig. \ref{fig4}, we show the $V$-dependence of $\lambda$ for the OTE ($s$-wave) and ESE ($d$-wave) at $U/t_1=4.0$ and $T/t_1=0.1$.
The value of $\lambda$ for the ESE decreases as $V$ increases.
This is due to the development of charge fluctuations $\chi_c$:
For the spin-singlet pairing, $\chi_c$ competes with $\chi_s$ since 
the sign of the coefficients of $\chi_s$ is opposite to that of $\chi_c$ in the effective interactions.
The first order of $V$, $V({\bm q})$ in Eq. (\ref{veffs}), also competes with $\chi_s$.
On the other hand, the value of $\lambda$ for the OTE slightly increases
since the effective interaction for the spin-triplet pairing is enhanced by $V$.

\begin{figure}[h]
\begin{center}
\includegraphics[height=6cm]{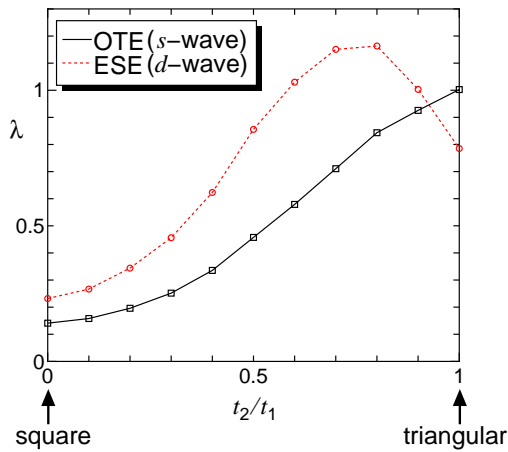}
\caption{$t_2$-dependence of $\lambda$ at $sf=0.92$ and $T/t_1=0.2$.}\label{fig3}
\end{center}
\end{figure}

In the above, we have shown that the odd-frequency pairing is favored near the
SDW state on a triangular lattice.
In the following, we clarify
how the above results are influenced by the lattice structure.
In Fig. \ref{fig3}, we show the $t_2$-dependence of $\lambda$ for OTE ($s$-wave) and ESE ($d_{x^2-y^2}$-wave),
where $t_2$ is a transfer integral on the diagonal.
This lattice corresponds to the square lattice in the case of $t_2/t_1=0$, and it corresponds to the triangular lattice in the case of $t_2/t_1=1$.
The critical point of the SDW state is different in each lattice, since the $\chi_0(q)$ value obtained is different.
Here, we introduce the Stoner factor, $sf=U\chi_{max}$, which gives the closeness to the SDW state,
where $\chi_{max}$ is the maximum of $\chi_0(q)$.
$sf=1$ corresponds to the SDW critical point, and $sf$ is less than unity in the paramagnetic phase.
We fixed the Stoner factor to be $sf=0.92$ for varying the values of $t_2/t_1$
for the purpose of keeping the distance from the SDW critical point.
The values of $\lambda$ for both OTE ($s$-wave) and ESE ($d_{x^2-y^2}$-wave) increase with $t_2$ for $t_2/t_1<0.7$,
since $U$, which is determined by $U=sf/\chi_{max}$, increases with $t_2$.
The value of $\lambda$ for ESE ($d_{x^2-y^2}$-wave) decreases with $t_2$ in 
the region $t_2/t_1>0.8$,
while the magnitude of $U$ used in the actual calculation increases 
with $t_2$.  
This is because geometrical frustration becomes 
significant with increasing $t_2$ and the resulting nesting condition becomes worse.
On the other hand, $\lambda$ for OTE ($s$-wave) continues to increase with $t_2$ in the region $t_2/t_1>0.8$,
since the frustration does not affect the instability of $s$-wave pairings.

\begin{figure}[h]
\begin{center}
\includegraphics[height=6cm]{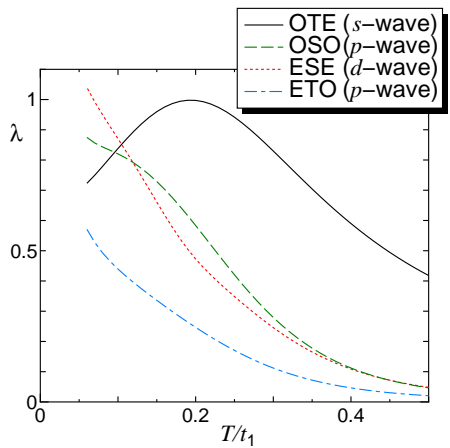}
\caption{$T$-dependence of $\lambda$ for $U/t_1=2.88$, $V=0$, $t_2/t_1=-0.6$.}\label{fig5}
\end{center}
\end{figure}

Finally, we show that odd-frequency pairing can be realized in a
frustrated square lattice with next nearest neighbor hopping
$(t_2)$, where the band dispersion is given as $\varepsilon_{\bm k}=-2t_1(\cos(k_x)+\cos(k_x))-4t_2\cos(k_x)\cos(k_y)$.
In Fig. \ref{fig5}, the $T$-dependence of $\lambda$ is shown for $t_2/t_1=-0.6$, $U/t_1=2.88$.
It can be shown that OTE ($s$-wave) is dominant at high temperatures as in
the case of the triangular lattice.
The AF fluctuation is suppressed due to the geometrical
frustration induced by large $t_2$.
Therefore, the value of $\lambda$ for ESE $d_{x^2-y^2}$-wave pairing
decreases and becomes smaller
than that of the OTE pairing.
Note that $V$ enhances the value of $\lambda$ of the
$d_{xy}$-wave pairings in the square lattice
while it suppresses that of the $d_{x^2-y^2}$-wave pairing.
However, the value of $\lambda$ for the $d_{xy}$-wave
pairings remains smaller than that of the $d_{x^2-y^2}$-wave.
Thus, we suggest that odd-frequency pairing appears in geometrical
frustrated systems.

In conclusion, 
we have investigated whether odd-frequency pairing states can be realized 
in the extended Hubbard model at half-filling on a 2D triangular 
lattice using the RPA. 
The odd-frequency spin-triplet $s$-wave pairing is dominant for a wide 
temperature regime.
The instability of the spin-triplet $s$-wave state is enhanced by the off-site Coulomb interaction $V$
because the effective interaction for the triplet channel is enhanced by $V$.
On the other hand, since the effective interaction for the spin-singlet 
channel is reduced by $V$, 
the even-frequency spin-singlet $d$-wave state is suppressed.
It is an interesting future problem to 
study the realization of odd-frequency pairs 
in the presence of a magnetic field or modulation of the 
pair potential. 
In addition, 
clarifying the superconducting properties 
of odd-frequency superconductors is also an interesting issue. 
Since phase sensitive probes, $e.g.$, 
tunneling and Josephson effects, are crucial to identify the 
pairing symmetry \cite{Tsuei,TK} in unconventional 
superconductors, similar studies on odd-frequency superconductors 
will become important \cite{Linder,Fominov}.

One of the authors, Y. T., 
would like to thank M. Vojta, A.V. Balatsky, 
D. J.  Scalapino and M. Oshikawa for valuable discussions.


\begin{thebibliography}{99}


\bibitem{Tsuei}
C. C. Tsuei and J. R. Kirtley, Rev. Mod. Phys. {\bf 72} (2000) 969.

\bibitem{bere}
V. L. Berezinskii:
JETP Lett. {\bf 20} (1974) 287.

\bibitem{bala}
A. Balatsky and E. Abrahams:
Phys. Rev. B. {\bf 45} (1992) 13125; 
E. Abrahams,  
A. Balatsky, D. J. Scalapino and J. R.
Schrieffer, Phys. Rev. B \textbf{52} (1995) 1271.

\bibitem{Bulut}
N. Bulut, D. J. Scalapino, and S.R. White, Phys. Rev. B 
\textbf{47} (1993) 14599.  

\bibitem{Coleman}
P. Coleman, E. Miranda and A. Tsvelik, Phys. Rev.  B  \textbf{49} (1994) 8955; 
P. Coleman, A. Georges, and A. M. Tsvelik, J. Phys. Condens. Matter 
\textbf{9} (1997) 345; 
O. Zachar, S. A. Kievelson and V. J. Emery, 
Phys. Rev. Lett. {\bf 77} (1996) 1342. 

\bibitem{Vojta}
M. Vojta and E. Dagotto: Phys. Rev. B {\bf 59} (1999) R713.

\bibitem{Fuseya}
Y. Fuseya, H. Kohno and K. Miyake: J. Phys. Soc. Jpn. {\bf 72} (2003) 2914.

\bibitem{Zhang}
G. Q. Zheng, N. Yamaguchi, H. Kan, Y. Kitaoka, J. L. Sarrao, P. G. Pagliuso, N. O. Moreno and J. D. Thompson, 
Phys. Rev. B, \textbf{70} (2004) 014511;
S. Kawasaki, T. Mito, Y. Kawasaki, G.-q. Zheng, Y. Kitaoka, D. Aoki, 
Y. Haga, and Y. Onuki, Phys. Rev. Lett. \textbf{91} (2003) 137001.

\bibitem{Efetov1} F. S. Bergeret, A. F. Volkov, and K. B. Efetov, Phys. Rev.
Lett. \textbf{86} (2001) 4096

\bibitem{Efetov2}
F. S. Bergeret, A. F. Volkov, and K. B. Efetov, Rev. Mod.
Phys. \textbf{77} (2005) 1321.

\bibitem{triplet}
Y. Tanaka and A.A. Golubov: Phys. Rev. Lett. \textbf{98} (2007) 037003; 
Y. Asano, Y. Tanaka,  A. A. Golubov and S. Kashiwaya, 
Phys. Rev. Lett. \textbf{99}  (2007) 067005. 

\bibitem{inhomo}
Y. Tanaka, A.A. Golubov, S. Kashiwaya and M. Ueda:
Phys. Rev. Lett. {\textbf 99} (2007) 037005; 
M. Eschrig, 
T. Lofwander, Th. Champel, J.C. Cuevas and G. Schon,
J. Low Temp. Phys. \textbf{147} (2007) 457; 
Y. Tanaka, Y. Tanuma, and A. A. Golubov, 
Phys. Rev. B \textbf{76} 054522 (2007). 


\bibitem{Miyake}
K. Miyake, S. Schimitt-Rink and C.M. Varma, Phys. Rev. B {\bf 34}  433 (1986);
D. J. Scalapino, E. Loh, Jr. and J. E. Hirsh, 
Phys. Rev. B {\bf 34}  8190 (1986); 
N. E. Bickers, D. J. Scalapino and S. R. White,
Phys. Rev. Lett. {\bf 62}, 961 (1989);
H. Yokoyama, Y. Tanaka, M. Ogata and H.Tsuchiura,
J. Phys. Soc. Jpn. \textbf{73}, 1119 (2004).

\bibitem{Kondo}
H. Kino and H. Kontani, J. Phys. Soc. Jpn. {\bf 67}  3695 (1998); 
H. Kondo and T. Moriya, J. Phys. Soc. Jpn. {\bf 67}  3695 (1998); 
H. Kondo and T. Moriya, J. Phys. Soc. Jpn. {\bf 68}  3170 (1999).

\bibitem{TK}
M. Sigrist and T. M. Rice, Rev. Mod. Phys. {\bf 67} 503 (1995), 
D. J. Van Harlingen, Rev. Mod. Phys. {\bf 67} 515 (2995), 
Y. Tanaka and S. Kashiwaya, Phys. Rev. Lett. {\bf 74} 3451 (1995), 
S. Kashiwaya and Y. Tanaka, Rep. Prog. Phys. {\bf 63} 1641  (2000), 

\bibitem{Linder}
J. Linder, T. Yokoyama, Y. Tanaka, Y. Asano, and A. Sudbo,  
Phys. Rev. B, {\bf 77}, 174505 (2006);
J. Linder, T. Yokoyama, and A. Sudbo,  
Phys. Rev. B {\bf 77}, 174507 (2008)

\bibitem{Fominov}
 Ya. V. Fominov, JETP Letters 86, 732 (2007) [Pis'ma v ZhETF 86, 842 (2007)] 




\end{thebibliography}
\end{document}